\documentclass{article}
\usepackage[utf8]{inputenc}

\usepackage{comment}
\usepackage{amsmath}
\usepackage{graphicx}
\usepackage{authblk}
\usepackage{fullpage}
\usepackage[colorinlistoftodos]{todonotes}
\usepackage[dvipdfmx]{hyperref}
\usepackage{multirow}
\usepackage{soul}

\begin{document}

\title{Transport collapse in dynamically evolving networks}

\author[1,2]{Geoffroy Berthelot\href{https://orcid.org/0000-0003-4036-6114}}
\author[3,4]{Liubov Tupikina\href{https://orcid.org/0000-0002-7169-5706}}
\author[5]{Min-Yeong Kang\href{https://orcid.org/0000-0002-4441-514X}}
\author[6]{J\'{e}r\^{o}me Dedecker}
\author[7]{Denis Grebenkov\href{https://orcid.org/0000-0002-6273-9164}}

\affil[1]{Institut National du Sport, de l'Expertise et de la Performance (INSEP), 75012 Paris, France}
\affil[2]{Research Laboratory for Interdisciplinary Studies (RELAIS), 75012 Paris, France}
\affil[3]{The Center for Research and Interdisciplinarity, Paris, France 75004}
\affil[4]{Bell Labs Nokia, France}
\affil[5]{Medipixel, Inc., 04037 Seoul, South Korea}
\affil[6]{Universit\'e de Paris, France}
\affil[7]{Laboratoire de Physique de la Mati\`ere Condens\'ee (UMR 7643), CNRS -- Ecole Polytechnique, IP Paris, 91128 Palaiseau, France}

\providecommand{\keywords}[1]{\textbf{\textit{Keywords---}} #1}

\maketitle

\begin{abstract}
Transport in complex networks can describe a variety of natural and human-engineered processes including biological, societal and technological ones. However, how the properties of the source and drain nodes can affect transport subject to random failures, attacks or maintenance optimization in the network remain unknown. In this paper, the effects of both the distance between the source and drain nodes and of the degree of the source node on the time of transport collapse are studied in scale-free and lattice-based transport networks. These effects are numerically evaluated for two strategies, which employ either transport-based or random link removal. Scale-free networks with small distances are found to result in larger times of collapse. In lattice-based networks, both the dimension and boundary conditions are shown to have a major effect on the time of collapse. We also show that adding a direct link between the source and the drain increases the robustness of scale-free networks when subject to random link removals. Interestingly, the distribution of the times of collapse is then similar to the one of lattice-based networks.
\end{abstract}

\keywords{transport in complex systems, resistor grid, scale-free network}

\section{Introduction}
Transport in complex (or scale-free) networks is relevant for various natural and artificial systems \cite{Serov2015,Greb2005,Newman2004,Lambiotte}. These networks have a degree distribution that follows a power law $P(k) \sim k^{-\gamma}$ (or truncated power law) \cite{Barabasi,HavlinStanley}, where $k$ is the node degree and $\gamma$ is an exponent. One often considers resistor networks, in which the transport is modeled as a current (or flux) generated by a source and transiting throughout the links to a drain. Nodes are assigned with electric potentials, while each link has a current (or flux). When applying a difference of potential between the source and drain nodes, the transport self-organizes into a peculiar arborescent configuration, with a tree-like structure emerging from the source and another one converging to the drain \cite{ResisIpaper}. These two trees merge into a large cluster with evenly distributed potentials, the so-called Quasi-Equipotential Cluster (QEC) \cite{ResisIpaper, lopez2005anomalous}. The QEC concentrates low fluxes, while high fluxes are instead concentrated in the links originating from the source and terminating at the drain \cite{ResisIpaper}.

Natural and artificial systems forming such networks can be altered with time, and often exhibit properties related to the percolation phenomena, where links are lost or destroyed with time. For instance, protein bonds may be lost with time in proteins networks, while proteins, the nodes of the network, remain present \cite{kovacs2019network}. Other examples include the lung airway tree \cite{mauroy2004optimal}, braided streams \cite{connor2018let, rinaldo2014evolution}, social networks \cite{shao2015percolation} and transportation networks for passengers \cite{mattsson2015vulnerability, pan2021vulnerability, von2012tale, zhang2018review}. Previous studies shown that scale-free networks are fragile and prone to collapse. This was demonstrated theoretically \cite{holme2002attack} and for multiple dynamical real-world systems, such as (species) evolution and ecosystems \cite{bellwood2004confronting, hanel2008solution}, the financial sector \cite{haldane2011systemic} and social networks \cite{murase2015modeling}. The stability of such systems can be jeopardized with the removal of just one element, possibly leading to collapse \cite{rizzo2003sudden, csekerciouglu2004ecosystem, elliott2014financial, crucitti2004model, gao2013percolation}. One well-known topological feature that improves the robustness of the network is the existence of cycles, \textit{i.e.} directed closed paths whose only repeated vertices are the first and the last. They were shown to provide stability to systems such as biochemical networks for instance \cite{ma2009defining, reznik2010stability}.

We previously studied three different strategies of a resistor network evolution by progressive removal of the weakest, random or strongest link at each time step $t$. These strategies can model respectively intentional attacks on the network, its random failures, and progressive network optimizations when the weakest and thus least useful links are removed (the pseudo-Darwinian strategy). In all three cases, transport eventually collapses at a time $t_c$, which depends on the chosen evolution strategy and a set of network parameters. In particular, it was shown that low $\gamma$ values yield high $t_c$ values \cite{berthelot2020pseudo}. In fact, low $\gamma$ results in an increased proportion of hubs (nodes with a high degree), which in turn increases redundancy, i.e. the number of paths between each pair of nodes in the network \cite{o2015network}. This is a key element to improve network robustness and therefore to delay the time of collapse $t_c$. Based on our former simulations, the evolution strategy which produced the smallest $t_c$ values corresponded to the removal of the strongest links, i.e. the links which hold the largest fluxes. These links are directly connected to the source and drain nodes and removing such links results in a rapid disconnection of the source or the drain from the rest of the network and thus the transport collapse. On the opposite, largest values of $t_c$ were obtained using the pseudo-Darwinian strategy when removing links with the lowest fluxes. As stated above, such links are located in the QEC and thus contribute less to the transportation system \cite{berthelot2020pseudo}. Finally, values of $t_c$ for the random evolution usually take place between the two above strategies. As a consequence, the evolution strategies with the removal of the weakest and the strongest links can serve to identify lower and upper bounds for $t_c$ values respectively. However, beside the parameter $\gamma$, the other topological properties affecting $t_c$ remain unclear.

In this paper, we investigate how the degree of the source and the distance (in terms of nodes) between the source and drain nodes can affect $t_c$. These two properties are studied in the context of percolation, in order to understand what role they may play in delaying $t_c$.  We also investigate resistor grids with regular, lattice-based geometry, which are widely employed as models for many problems in science and engineering \cite{kirkpatrick1973percolation,klein1993resistance}. Our interest in including these networks lies in their homogeneous topology, which can serve as a reference for comparing the effects of the studied properties (distance and degree). We only investigate finite $D$-dimensional lattice-based networks, which can be constructed \cite{moody2015resistor} and used in stealth \cite{sudhendra2011novel}, sensors in robotics \cite{medina2016resistor} or for energy dissipation in road or railroad vehicles \cite{mirzaei2012retard,tabarra2005measurement} for instance. We also aim at studying how the dimensionality can increase the robustness of such networks which can, in turn, help in increasing circuit boards reliability.

\section{Methods}
In this section, we describe the construction and parameters of the networks, which can be either scale-free networks (section \ref{sec:Network structure and transport}) or lattice-based networks (section \ref{sec:Comparison with lattice-based networks}). Each network (or graph \cite{iniguez2020bridging}) is undirected and contains a total of $N_0$ nodes and $L_0$ links. The transport in these networks is modeled using two main elements (section \ref{sec:Network structure and transport}). First, two nodes are selected: one for the source and the other for the drain, such that the transport starts from the source node and ends at the drain node. The selection of the source and drain nodes can either be random or deterministic, depending on the effect we aim to assess (see section \ref{sec:Network structure and transport} for a random selection and section \ref{sec:Delta} for a specific selection of the source and drain pair). Second, the transport mechanism is described by the Kirchhoff's equations, i.e. the sum of currents entering any node is equal to the sum of currents leaving it. This basic rule ensures the conservation of charge in electric circuits or the conservation of mass in transport systems. An application of the electrical potential differences (voltage) between the source and the drain nodes results in a direction of the current from the source to the drain. In other words, the flux self-organizes from the source to the drain \cite{ResisIpaper}. This setup can serve as an analogy to transport for several real world phenomena (see references in the introduction). We then detail two strategies of network evolution, in which we progressively remove links at each time steps depending on the chosen strategy (section \ref{sec:Network evolution}). The evolution stops when either the transport is no more possible between the source and drain, or when a portion of the network is separated from the rest. These conditions define the time of collapse, which depends on the initial network, the choice of the source and drain, and the network evolution. If at least one of these elements is random (e.g., the random evolution strategy), the time of collapse is a random variable, and we are interested in characterizing its probability density via numerical simulations.

\subsection{Network structure and transport}
\label{sec:Network structure and transport}
\begin{figure}
\centering
\includegraphics[width=1\textwidth]{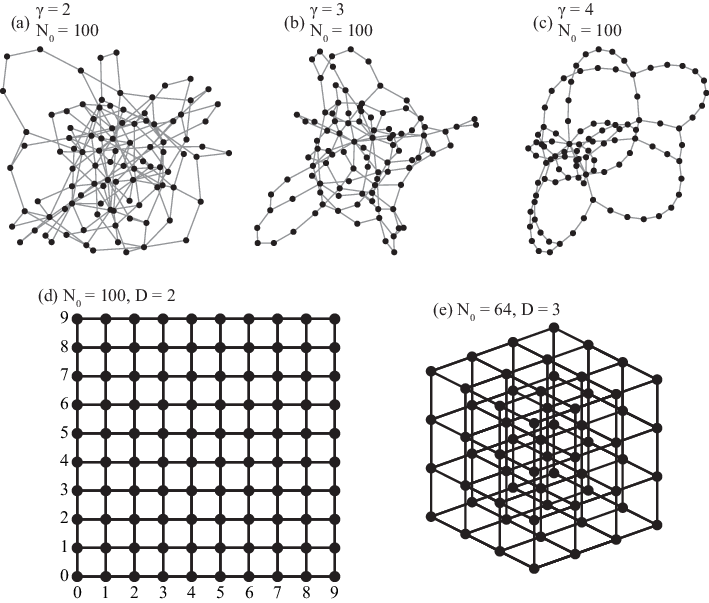}
\caption{Examples of network realizations. Scale-free networks with $\gamma=\{2,3,4\}$ and $N_0=100$ are illustrated in panels (\textbf{a}) - (\textbf{c}). Grid networks with reflecting boundary conditions and dimensions $D=\{2,3\}$ and sizes $N_0=\{100,64\}$ are presented in panels (\textbf{d}), (\textbf{e}).}
\label{fig:Fig1_Illustrations}
\end{figure}

We construct a random scale-free network with $N_0$ nodes using the uncorrelated configuration model with a given degree exponent $\gamma$ \cite{Satoras} (Fig. \ref{fig:Fig1_Illustrations}). In each realization of the network with a prescribed exponent $\gamma$, we select a pair of source and drain nodes, at which the potential is fixed to be 1 and 0, respectively. In some cases, this selection is performed randomly (uniformly) among all pairs of nodes. In other cases, we run simulations for all pairs of nodes. The rule of the selection will be specified for each study. As a permutation of the source and the drain in any pair does not change the (absolute values of) fluxes and potentials (see below), such two choices lead to the same evolution and result in the same time of collapse. For this reason, one can reduce the total number of pairs of distinguishable nodes, $N_0(N_0-1)$, to the twice smaller number of pairs of indistinguishable nodes, $N_0(N_0-1)/2$. The flux through the network satisfies conservation of mass \cite{Nicolaides}: at every node $i$ we impose $\sum_{j} q_{i,j} = 0$, where $q_{ij}$ is the flux through the link connecting nodes $i$ and $j$. These fluxes are calculated as:
\begin{equation}\label{flux_eq}
    q_{i,j} = - \left(P_i - P_j\right)
\end{equation}
where $P_i$ and $P_j$ are the potentials at the nodes $i$ and $j$ respectively, and we set the unit resistance for all links. Note that we have studied similar networks with distance-dependent resistances and showed their marginal effect on the transport \cite{ResisIpaper,berthelot2020pseudo}. In such a setup, the system of linear Kirchhoff\textquoteright s equations \cite{redner2001guide} describing the transport in a network is:
\begin{equation}\label{Kirchhoff_sys_raw}
    -\mathbf{L}'\mathbf{P}'^\top = 0
\end{equation}
with $\top$ denoting transposition, $\mathbf{L}'$ the Laplacian matrix of $N_0 \times N_0$ elements and $\mathbf{P}'^\top$ the $N_0 \times 1$ column vector of known (fixed) and unknown potentials. Each element of $\mathbf{L}$ is defined as:
\begin{equation}\label{Laplacian_matrix}
        \mathbf{L}_{i,j} =
        \begin{cases}
            k_i & \text{if $i = j$}\\
            -1 & \text{if node $i \neq j$, and $i$ is adjacent to node $j$}\\
            0 & \text{otherwise}
        \end{cases}
\end{equation}
After removing the source and drain nodes, at which the potentials are fixed, the above system can be rewritten in the following form:
\begin{equation}\label{Kirchhoff_sys}
    \mathbf{L}\mathbf{P}^\top = \mathbf{S}^\top
\end{equation}
where $\mathbf{P}^\top$ is the vector of $N_0-2$ unknown potentials, $\mathbf{L}$ is the Laplacian matrix of $(N_0-2) \times (N_0-2)$ elements obtained from $\mathbf{L}'$ after removal of two lines and two columns corresponding to the source and drain nodes, and $\mathbf{S}^\top$ is a $(N_0-2) \times 1$ column vector, in which each element corresponds to the total flux exiting each node $i$:
\begin{equation}\label{S_vector}
        \mathbf{S} =
        \begin{cases}
            1 & \text{if node $i$ is adjacent to the source node}\\
            0 & \text{otherwise}
        \end{cases}
\end{equation}
The system described in eq. (\ref{Kirchhoff_sys}) is solved for $\mathbf{P}$ numerically using a custom routine in Matlab. The distributions of potentials on nodes and fluxes in links are then obtained. In particular, we compute the total flux $Q$, that is the sum of fluxes exiting from the source node.

Let us come back to the above statement that the exchange of roles between the source and the drain does not change the total flux. This statement follows directly from the linearity of the Kirchhoff\textquoteright s equations and zero flux condition at each node. In fact, if $\mathbf{P}'$ is a solution of Eq. (\ref{Kirchhoff_sys_raw}), then $\mathbf{P}'' = \mathbf{I} - \mathbf{P}'$ is also a solution because $\mathbf{L}' \mathbf{I} = 0$, where $\mathbf{I} = (1,1,\ldots,1)^\top$. But the removal of the source and drain nodes from $\mathbf{P}''$ is equivalent to permuting the source and the drain in the corresponding vector $\mathbf{S}$ in Eq. (\ref{Kirchhoff_sys}). In other words, one can first shift the potential by $-1$ (i.e., the source has the potential $0$ instead of $1$ and the drain has potential $-1$ instead of $0$) and then multiply them by $-1$.  This is of course consistent with the fact that an electric potential in physics is defined up to a constant. This property can have practical implications for real life networks that can also be described by resistor grids, such as supply chain networks, water distribution systems, road/air transportation networks, electric power networks and social networks for instance \cite{zhao2007optimal, allesina2010performance, yazdani2011resilience, rosas2007topological, nardelli2014models}.

\subsection{Network evolution}
\label{sec:Network evolution}
The constructed network evolves at discrete steps according to a pre-selected strategy. At each evolution step, we solve the system of Kirchhoff's equations (\ref{Kirchhoff_sys}) and remove either the weakest link (pseudo-Darwinian strategy) or a random link (random strategy) \cite{berthelot2020pseudo}. After a link removal, we also remove ``dead-end'' nodes (i.e., nodes whose degree equals $1$), thus requiring any existing node after an evolutionary step to have at least degree $2$. Each evolution step is associated with ``time'' $t$, and $t_0=0$ being the initial time, when the evolution starts. We denote by $Q_0$ and $L_0$ the total flux and the number of links respectively at $t_0$.

We aim at estimating the time of collapse $t_c$ of the network. The evolution ends when at least one of the following conditions is met: (\textit{i}) no path exists between the source and the drain (i.e. the source and drain are disconnected), (\textit{ii}) the source or the drain is removed from the network, (\textit{iii}) a portion of the network -- a subgraph containing more than one node -- is disconnected from the rest of the network. This last condition is implemented to reflect natural systems such as energy, transportation or biological networks, where it is not desirable to remove a portion of nodes from the rest of the network. The end of the simulation defines $t_c$, corresponding to a situation where the transport can no longer be maintained through the whole network.

We generally obtain a distribution of $t_c$ by running many simulations for a given set of network parameters. The total number of simulations is given as $n_{\rm r} \cdot n_{\rm p} \cdot n_{\rm s}$, where $n_{\rm r}$ is the number of random realizations of the network, $n_{\rm p}$ is the number of choices of the source-drain pair (either $n_{\rm p} = 1$ for a single random choice or $n_{\rm p} = N_0(N_0-1)/2$) for a systematic analysis of all pairs, and $n_{\rm s}$ is the number of random evolutions (note that $n_{\rm s} = 1$ for the pseudo-Darwinian strategy, which is deterministic).

\subsection{Comparison with lattice-based networks}
\label{sec:Comparison with lattice-based networks}
For comparison, we also investigate the transport in lattice-based networks of dimension $D=\{2,3,4\}$. Such a network consists of integer points on $\{1,\ldots,N\}^D$, each of them being connected to its nearest neighbors (Fig. \ref{fig:Fig1_Illustrations}). We consider two boundary conditions: periodic boundary condition (when each node degree $k$ is equal to $2D$) and reflecting (also known as ``open'') boundary condition (with $k \in [D,2D]$). We aim here to reveal the role of topological properties of the network, which are different between lattice-based and scale-free networks. In particular, we assess whether lattice-based networks are more robust than scale-free ones for both evolution strategies. Another point of interest is to assess the effect of dimensionality and of the type of boundary conditions on $t_c$. To reduce finite-size effects, we normalize each $t_c$ by $L_0$.

\section{Results}
\subsection{Effect of the distance between source and drain}
\label{sec:Delta}

\begin{figure}
\centering
\includegraphics[width=1\textwidth]{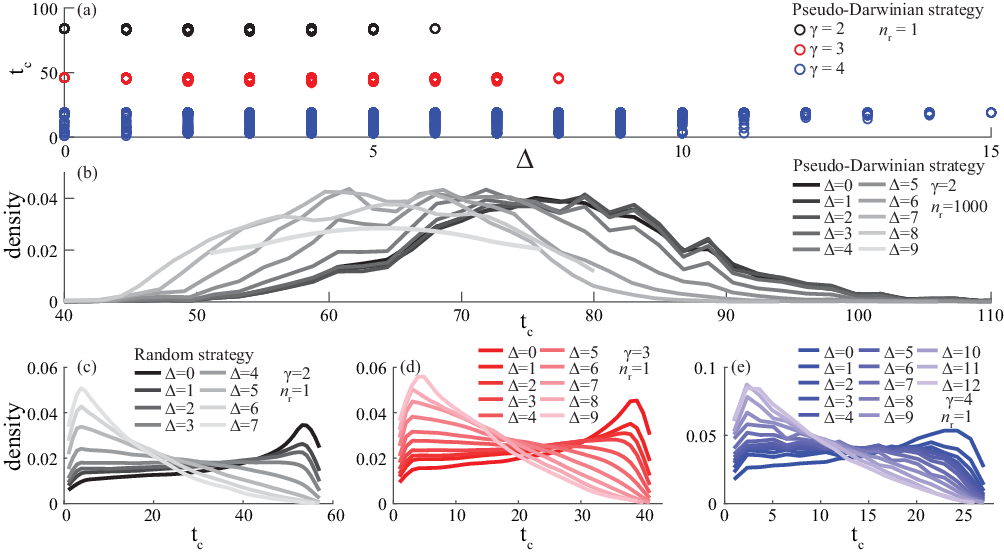}
\caption{Effect of distance $\Delta$ between the source and the drain on the distribution of the time of collapse $t_c$ for scale-free networks of
size $N_0 = 100$. (\textbf{a}) $t_c$ as a function of $\Delta$ for a single network realization ($n_{\rm r} = 1$) for each of three values of $\gamma = \{2,3,4\}$ and the pseudo-Darwinian strategy ($n_{\rm s} = 1$). (\textbf{b}) The probability density of $t_c$ for $n_{\text{r}}=1000$ realizations, $\gamma = 2$ and the random strategy ($n_{\rm s} = 1$). (\textbf{c} - \textbf{e}) The probability density of $t_c$ for one realization ($n_{\text{r}}=1$) and $n_{\rm s} = 1000$ simulations for the random strategy with different values of $\gamma$: $\gamma = 2$ (\textbf{c}), $\gamma = 3$ (\textbf{d}), and $\gamma = 4$ (\textbf{e}). Probability densities in panels (\textbf{b} - \textbf{e}) are smoothed using an Epanechnikov kernel.}
\label{fig:Fig2_Delta}
\end{figure}

We start by studying the effect of the distance $\Delta$ (measured in nodes) between the source and the drain on the time of collapse $t_c$ of a scale-free network for a given $\gamma$ value. Note that $\Delta=0$ means that there is a direct link between the source and the drain. Figure \ref{fig:Fig2_Delta} summarizes our results for a scale-free network of size $N_0 = 100$.

First, we produce a single (random) realization of the scale-free network ($n_{\text{r}} = 1$) and consider all possible pairs of source and drain nodes. For each pair, we run a pseudo-Darwinian strategy and compute the time of collapse $t_c$. This simulation yields $n_{\rm p} = N_0(N_0-1)/2 = 4950$ values of $t_c$, which are grouped according to the distance $\Delta$ (see panel (\textbf{a})). As the pseudo-Darwinian strategy is deterministic, different values of $t_c$ can be interpreted, to some extent, as the result of a ``random choice'' of the source and drain nodes. For $\gamma = \{2,3\}$, the values of $t_c$ for each $\Delta$ are almost identical, with a minor variation. In other words, the time of collapse is mainly determined by the distance $\Delta$ between the source and the drain and is almost independent of their particular location in the network. Due to a finite-size effect ($N_0 = 100$), the maximal value of $\Delta$ is limited to $6$ for $\gamma = 2$ and to $8$ for $\gamma = 3$. The behavior is different for $\gamma = 4$. One sees a much bigger variation of $t_c$ for each value of $\Delta$, meaning that the distance $\Delta$ alone does not determine the time of collapse, and the location of the source and drain nodes does matter here. This qualitative difference originates from the structure of the scale-free network, which has a larger proportion of ``hubs'' for smaller $\gamma$.

The above results were obtained for a particular random realization of the network. To study the effect of stochasticity (i.e. node degree and wiring), we generated $n_{\text{r}} = 1000$ random realizations of the scale-free network with $N_0 = 100$ and $\gamma = 2$. For each realization, we consider all possible pairs of the source and drain nodes and for each assignment, we perform the pseudo-Darwinian evolution to determine the time of collapse. In this way, we obtain $n_{\text{r}} N_0(N_0-1)/2$ values of $t_c$. We split them into groups according to the distance $\Delta$ and then plot a probability density of $t_c$ (a rescaled empirical histogram) for each group on panel (\textbf{b}). One sees that these histograms are still relatively narrow (with the standard deviation $\sim 10$ being much smaller than the mean) but exhibit a larger variation than for a single realization of the network. It is expected that randomness of the network structure broads the distribution of the time of collapse. As $\Delta$ increases, the maximum of the probability density shifts to the left (to smaller $t_c$) suggesting that the network becomes less robust. In other words, large $\Delta$ is generally associated with small $t_c$.

It is instructive to check whether this statement remains valid for the random strategy. For this purpose, we generated a single realization of the network and performed $n_{\rm s} = 1000$ random evolutions of this network until its collapse. As previously, we consider all pairs of the source and drain nodes. In this way, we generated $1000 N_0(N_0-1)/2$ values of $t_c$ that were split into groups according to the distance $\Delta$.  The probability densities of $t_c$ for each group are shown on panels (\textbf{c}), (\textbf{d}), and (\textbf{e}) for $\gamma = 2,3,4$, respectively.
One observes a similar trend that the maximum of the distribution is shifted to the left as $\Delta$ increases. In turn, these distributions are more skewed than that of panel (\textbf{b}) for the pseudo-Darwinian strategy.

\subsection{Effect of artificial connection}
\begin{figure}
\centering
\includegraphics[width=0.6\textwidth]{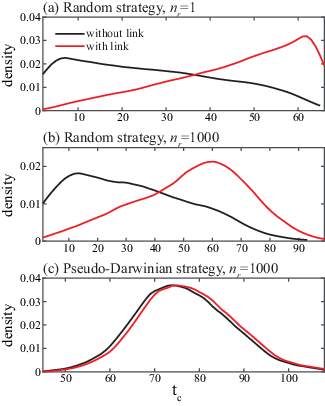}
\caption{Effect of adding a direct link between the source and the drain for a scale-free network with $N_0 = 100$, $\gamma=2$ and a single choice of the source-drain pair ($n_{\rm p} = 1$) under the constraint $\Delta \geq 4$.  In panel (\textbf{a}), the probability density of $t_c$ is shown for the random strategy and a single realization ($n_{\text{r}}=1$).  A total of $n_{\rm s} = 2\times10^4$ evolutions are performed for each of the two constructions (with and without the direct link).  Similarly in (\textbf{b}), the probability density of $t_c$ is given for the same strategy, but using $n_{\text{r}}=10^3$ realizations and $n_{\rm s} = 10^4$ simulations for each realization. In panel (\textbf{c}), the distribution for $t_c$ is given for the pseudo-Darwinian strategy, using one evolution per realization ($n_{\rm s} = 1$) and $n_r=10^3$ realizations of the network for each of the two constructions. Probability densities are smoothed using an Epanechnikov kernel.}
\label{fig:Fig3_shunt}
\end{figure}

Since the distance $\Delta$ strongly affects the time of collapse, one may wonder how an addition of a direct link between the source and the drain can change $t_c$ for two strategies. The direct link resets the distance $\Delta$ of an already existing network to $0$. For this purpose, we fix $N_0=100$ amd $\gamma=2$, impose $\Delta \geq 4$ and compare two empirical distributions of $t_c$, without and with an artificial connection. To fulfill the condition $\Delta \geq 4$, we randomly select a pair of nodes and check whether their distance is greater or equal to $4$; if yes, these nodes are assigned to be the source and the drain; otherwise, a new pair is selected, and so on, until the condition is satisfied.
Figure \ref{fig:Fig3_shunt} summarizes the results for two strategies in three settings.

(i) We start with a single random realization of a scale-free network ($n_{\rm r} = 1$), which is independently evolved $n_{\rm s} = 2\times10^4$ times by random strategy to get an empirical distribution of $t_c$. Then, a link between the source and the drain is added to the initially constructed network, such that $\Delta$ now equals 0, and $2\times10^4$ evolutions are performed again to get another distribution of $t_c$. Panel (\textbf{a}) compares two empirical distributions. One sees that adding a single link between the source and drain nodes delays the time of collapse.  Moreover, it produces an important change in the distribution of $t_c$, switching the mode of the distribution from low to high values. Interestingly, the modified distribution looks similar to the distribution for lattice-based networks (Fig. \ref{fig:Fig5_grids}), whose properties are controlled by dimensionality $D$ and boundary conditions (see below).

(ii) To study the effect of stochasticity, we then constructed $n_r=10^3$ realizations of the scale-free network, and each of these networks is evolved $n_{\rm s} = 10^4$ times by the random strategy to get the distribution of $t_c$. Another distribution is obtained after connecting the source and the drain for each initial network and evolving it $10^4$ times again. Panel (\textbf{b}) compares two distributions, which are similar to those shown in panel (\textbf{a}). We conclude that stochasticity of the network does not matter here.

(iii) Finally, we perform simulations similar to (ii), but the random strategy is replaced by the pseudo-Darwinian strategy. One can see on panel (\textbf{c}) that two distributions of $t_c$ are almost identical, i.e., the effect of an artificial connection is minor here. In fact, the inclusion of the direct link just requires one supplementary step to disconnect the source and the drain, thus replacing $t_c$ by $t_c+1$. This is not surprising because the direct link supports the strongest current and is thus removed at the very end of the pseudo-Darwinian evolution.

\subsection{Effect of the degree of the source}
\begin{figure}
\centering
\includegraphics[width=1\textwidth]{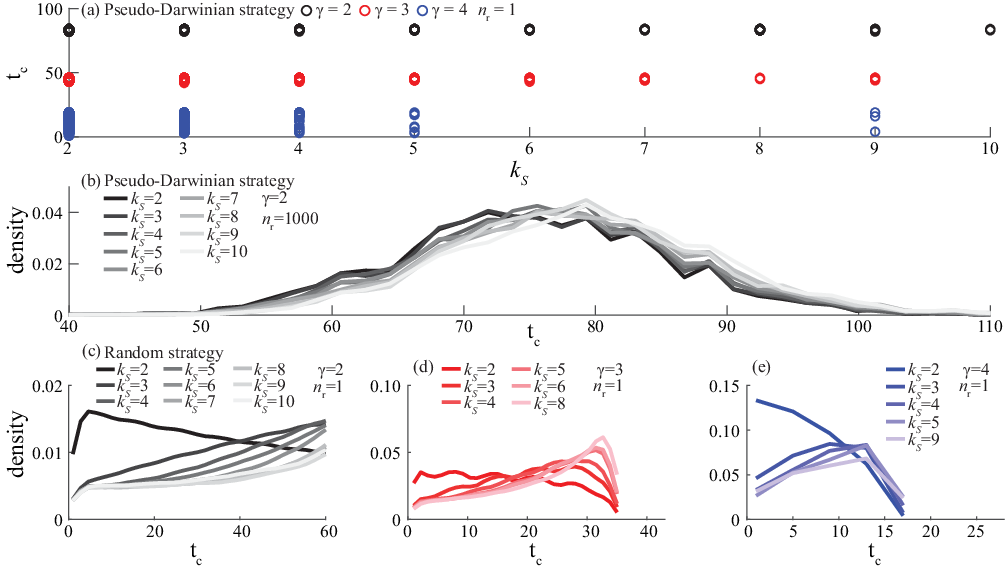}
\caption{Effect of the degree $k_S$ of the source on the distribution of $t_c$ for scale-free networks with size $N_0 = 100$. Panel (\textbf{a}):
$t_c$ as a function of $k_S$ for the pseudo-Darwinian strategy applied to a single random realization of the network ($n_{\rm r} = 1$) for each value of $\gamma$. Panel (\textbf{b}): the probability density of $t_c$ for the pseudo-Darwinian strategy applied to $n_{\text{r}}=1000$ random realizations of the scale-free network with $\gamma = 2$. Panels (\textbf{c} - \textbf{e}): the probability density of $t_c$ for $n_{\rm s} = 1000$ random strategy evolutions of a single random realization ($n_{\text{r}}=1$) of the network with different values of $\gamma$: $\gamma = 2$ (\textbf{c}), $\gamma = 3$ (\textbf{d}), and $\gamma = 4$ (\textbf{e}). Probability densities in panels (\textbf{b} - \textbf{e}) are smoothed using an Epanechnikov kernel.}
\label{fig:Fig4_source}
\end{figure}

The degree $k_S$ of the source node is another quantity that may affect the distribution of $t_c$. To investigate its role, we undertook a similar procedure as in Sec. \ref{sec:Delta}, namely, we first considered the pseudo-Darwinian strategy for a single realization of the scale-free network, then investigated the effect of stochasticity, and finally compared to the evolution by random strategy (Fig. \ref{fig:Fig4_source}). Panels (\textbf{a}) and (\textbf{b}) show that the distribution of $t_c$ is almost independent of the degree of the source when using the pseudo-Darwinian strategy. On the opposite, the random strategy results in an increase of $t_c$ when $k_S$ is greater than 2, see Fig. \ref{fig:Fig4_source}~(\textbf{c})-(\textbf{e}). The same results are obtained for the degree of the drain (not shown).

\subsection{Comparison with lattice-based networks}
\begin{figure}
\centering
\includegraphics[width=1\textwidth]{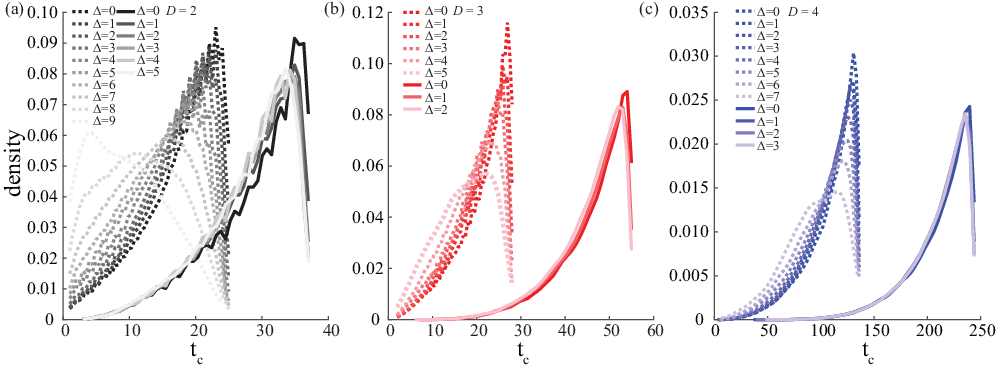}
\caption{Effect of the distance $\Delta$ between the source and the drain on the time of collapse $t_c$ for lattice-based networks of various dimensions $D=\{2,3,4\}$ that evolve via the random strategy. The shown probability densities of $t_c$ are obtained from $n_{\rm s} = 1000$ random evolutions for both open boundary conditions (dotted lines) and periodic ones (solid lines). Probability densities are smoothed using an Epanechnikov kernel.}
\label{fig:Fig5_grids}
\end{figure}

Let us now analyze how the topological structure of the network may affect the time of collapse. For this purpose, we compare the previous results for scale-free networks to those obtained for lattice-based networks. We fix the lattice dimension $D$, its size $N_0 = N^D$, and the type of boundary conditions (periodic or open). For each pair of the source and the drain (i.e., $n_{\rm p} = N_0(N_0-1)/2$), we run $n_{\rm s} = 1000$ evolutions by the random strategy to obtain the distribution of $t_c$ (Fig. \ref{fig:Fig5_grids}). Contrarily to the case of scale-free networks (see Fig. \ref{fig:Fig2_Delta}), there is only a little effect of $\Delta$ on the distribution of $t_c$ in all considered lattice-based networks. This effect is stronger for large $\Delta$ but this result may originate from finite-size effects (a limited number of configurations with large $\Delta$). Such networks have a homogeneous structure that can serve as a benchmark for comparison to other networks with a different topology. We also observe that the mode value of $t_c / L_0$ is affected by both the dimension and the boundary condition: (i) the higher dimension generally yields larger $t_c$; (ii) periodic boundary condition yield larger $t_c$, i.e., more robust networks.

In the case of deterministic the pseudo-Darwinian strategy, the only step that might allow for variability of $t_c$ is the choice of the source and drain nodes. Table \ref{Tab1} shows that this choice has almost no effect onto the time of collapse. For example, for the lattice-based network with $D=2$, $N_0=36$ and open boundary condition, there are only two values of $t_c / L_0$: $0.400$ (obtained for 4 source-drain pairs) and $0.417$ (obtained for the remaining 626 source-drain pairs, among $N_0(N_0-1)/2 = 630$ possible pairs). As in addition these two values are very close, we conclude that the time of collapse does not depend on the choice of the source and the drain, nor on their distance. This is in sharp contrast to the case of scale-free networks. In turn, $t_c$ depends on the dimension, the network size, and the type of boundary conditions.

\begin{table}[htbp]
\centering
\begin{tabular}{|c|c|c|c|c|c|}
        \hline
        $D$ & $N_0$ & $N_0(N_0-1)/2$ & Type of condition & $t_c / L_0$ & frequency\\
        \hline
        \multirow{4}{*}{2} & \multirow{4}{*}{36} & \multirow{4}{*}{630} & \multirow{2}{*}{open} & 0.400 & 4\\
         &  &  &  & 0.417 & 626\\  \cline{4-6}
         &  &  & \multirow{2}{*}{periodic} & 0.500 & 5\\
         &  &  &  & 0.514 & 625\\  \hline
        \multirow{4}{*}{3} & \multirow{4}{*}{125} & \multirow{4}{*}{7750} & \multirow{2}{*}{open} & 0.583 & 57\\
         &  &  &  & 0.587 & 7693\\  \cline{4-6}
         &  &  & \multirow{2}{*}{periodic} & 0.667 & 317\\
         &  &  &  & 0.669 & 7433\\  \hline
        \multirow{4}{*}{4} & \multirow{4}{*}{81} & \multirow{4}{*}{3240} & \multirow{2}{*}{open} & 0.625 & 8\\
         &  &  &  & 0.630 & 3232\\  \cline{4-6}
         &  &  & \multirow{2}{*}{periodic} & 0.750 & 21\\
         &  &  &  & 0.753 & 3219\\
        \hline
\end{tabular}
\caption{
Effect of dimension and periodic-vs-open boundary conditions of lattice-based networks on $t_c/L_0$ when using the pseudo-Darwinian strategy. There are two values for each dimension and type of boundary condition, depending on the location of the source and drain nodes. The frequency of the corresponding $t_c/L_0$ is also given.}
\label{Tab1}
\end{table}

\section{Discussion}
In this paper, we studied two evolution strategies that alter the network iteratively \cite{berthelot2020pseudo}. The pseudo-Darwinian strategy, which is deterministic and controlled by fluxes, results in targeted, transport-based evolution; in particular, it can represent the process of transport network optimization, in which least used elements are progressively removed. In turn, the random strategy is a stochastic procedure that acts independently of the fluxes and results in random, topologically-based evolution; it can model spontaneous failures and progressive degradation due to, e.g., aging of the network elements.  However, the evolution is driven by two transport-based mechanisms that are independent of the chosen strategy. First, the evolution is stopped (at the time of collapse $t_c$) when transport is no longer possible in the whole network, either because the source or the drain was removed, or because a portion of the network became disconnected (isolated). Second, when a link is removed from a node with degree $k=2$, the node becomes a ``dead-end'' ($k=1$) and no longer contributes to the transport in the rest of the network; it is then also removed. These two mechanisms imply that, independently of the chosen strategy, the outcomes remain transport-based.

In both strategies, we observed that $t_c$ is higher when the distance $\Delta$ is small, such as the network is more resilient to transport-based and topological-based evolutions when the source and the drain are close to each other. This is valid only for scale-free networks, as $\Delta$ has little effect in lattice-based networks (Fig. \ref{fig:Fig5_grids}). In scale-free networks, this topological effect was further demonstrated when adding a direct link between the source and the drain (Fig. \ref{fig:Fig3_shunt}), i.e., by creating one shortest path of length zero (1 link). Such an operation is equivalent to shunts in electronics where small devices are used for creating alternatives paths for electric currents \cite{graf1999modern}.  In biological mass-transfer networks, such as lung airways and blood vessels, these alternative direct paths arise naturally \cite{gompelmann2013collateral, santulli2013angiogenesis} or can be added deliberately to compensate insufficient flux due to large resistance in the system \cite{alexander2016coronary}. In our simulations, this shunt greatly changed the distribution of $t_c$ for the random strategy (Fig. \ref{fig:Fig3_shunt} (a) and (b)). On the opposite, this artificial modification had little to no effect when using the pseudo-Darwinian strategy (Fig. \ref{fig:Fig3_shunt} (c)). This difference is interesting because a small distance $\Delta$ tends to produce a larger $t_c$ for both strategies (Fig. \ref{fig:Fig2_Delta}), but the posterior topological modification (which results in $\Delta=0$) is only beneficial when using the random strategy (Fig. \ref{fig:Fig3_shunt}). It suggests that $\Delta$ is an indicator of robustness rather than the cause when it comes to transport-based collapses. Thus, the increased robustness not only depends on one-link connection between the source and drain nodes, but rather originates from the construction itself, implying a peculiar organization of links with small $\Delta$. Another interesting result of this artificial modification is that the collapse of transport is similar in scale-free and lattice-based networks (Fig. \ref{fig:Fig3_shunt} and \ref{fig:Fig5_grids}).

A similar observation can be made about the degree $k_S$ of the source node: while $k_S$ has an effect on $t_c$ when using the random strategy (Fig. \ref{fig:Fig4_source}~(c)-(e)), it plays very little (or no) role when using the pseudo-Darwinian strategy (Fig. \ref{fig:Fig4_source}~(b)).  This further highlights the fact that the two strategies differ in their behavior. As such, the degree of the source (or of the drain) cannot serve as a common indicator of robustness for both strategies. On the contrary, the distance $\Delta$ is more appropriate.

Several former studies showed that network integrity depends on a particular topology when subject to targeted or random attacks \cite{holme2002attack, albert2000error}. In particular, Holme {\it et al.} showed that Erd\H{o}s-R\'{e}nyi random networks are the most robust while scale-free networks are the most vulnerable to attacks on links or nodes \cite{holme2002attack}. In the present work, we compared the time of collapse of scale-free and lattice-based networks. In lattice-based networks, $t_c$ is related to the type of boundary conditions and the dimensionality $D$, with periodic boundary conditions and higher values of $D$ providing higher $t_c$ (Fig. \ref{fig:Fig5_grids}). The topology of lattice-based networks is similar to the one of hypercubic graphs $Q_n$, but with an increased number of nodes per edge. We stress that $Q_n$ are commonly used and appreciated in data sciences for both their versatile topology and efficiency in distributing numerical data between nodes in interconnection networks \cite{ostrouchov1987parallel, sardar2021novel, di2014traffic}. They are efficient structures for transport, which are also resilient to attacks \cite{9333978, di2014traffic}. This is in line with our results, which suggest that lattice-based networks of higher dimensions can lead to efficient transport and possibly higher resistance to topological attacks (Fig. \ref{fig:Fig5_grids}). This may also apply to scale-free networks, as suggested by Wu {\it et al.}, who studied the correlation between dimension and robustness of scale-free networks \cite{doi:10.1142/S0218348X19500671}. However, Wu {\it et al.} did not directly investigate link percolation in an evolutionary framework but rather relied on typical robustness metrics such as network efficiency and average edge betweenness, which can be incomplete in regards to our methodological framework. Yet, it remains to be seen if the dimension and fractal dimension can correlate with the time of collapse $t_c$. This problem will be investigated in a subsequent work.

Multiple former studies also demonstrated the usefulness of centrality measures for assessing node importance in terms of topology only \cite{iyer2013attack, lu2016vital, oldham2019consistency}. However, these studies were focused on the progressive removal of links or nodes under various strategies but without taking into account transport, which is critically relevant for various physical, natural and artificial systems. In particular, these studied did not take into account the two transport-based mechanisms discussed above. We demonstrate that introducing a single link between the source and the drain can increase the robustness of the network (Fig. \ref{fig:Fig3_shunt}). This artificial modification results in an increase of the (node-) betweenness centrality (BC) of both the source and drain nodes.  The BC is a practical computational metric for measuring network robustness subject to attacks \cite{FREEMAN1978215}. This metric essentially indicates the number of shortest paths going through a node. The BC also explains why dimensionality and periodic conditions are important for robustness in lattice-based networks as the BC depends on both. However, in our setup, it is difficult to provide a straight-forward interpretation of what BC really is. From a topological point of view, we study the evolution of a single network that remains to be a connected undirected graph at each time step. As such, one can typically consider the shortest paths that go through the source and drain nodes and simply ignore the transport information (flow). From the transport point of view, however, the network self-organizes and becomes a directed graph, with a flow originating from the source and ending at the drain \cite{ResisIpaper}. One can then take into account this flow and use it as the direction for the links, and accordingly compute the BC in this directed graph. However, this would result in the source and drain nodes having their BC values equal to 0, as the current originates from the source and ends up to the drain, resulting in these two nodes having no paths going through. These are distinct ``end-nodes'' of major importance in terms of the transport that cannot be ignored. Thus, typical metrics such as the BC or closeness centrality should be further updated with the transport information in order to better measure node importance in such a setup. For instance, one can potentially exploit new metrics such as the current flow betweenness centrality \cite{brandes2005centrality, newman2005measure, agryzkov2019variant}, which is based on random walks and employs the established connections between electric current flows and random walks \cite{doyle1984random}.

In summary, this work contributes to the understanding of transport in scale-free or lattice networks, and more generally, to dynamical real-world systems that can experience the progressive removal of links. This includes both natural and artificial networks, or networked systems: species evolution \cite{jackson2001historical, bellwood2004confronting, hanel2008solution}, biological systems \cite{dorogovtsev2003evolution}, such as the protein-interaction network \cite{jeong2001lethality, barabasi2004network}, metabolic network \cite{jeong2000large}, and cellular network \cite{Albert2001}, economic systems \cite{10.1371/journal.pone.0038924}, the financial sector \cite{haldane2011systemic, doi:10.1080/14697680400020325}, social networks \cite{murase2015modeling}, such as the author-collaboration network in social systems \cite{doi:10.1073/pnas.98.2.404, Albert2001}, communication systems such as the Internet \cite{10.1145/316194.316229, Albert2001, dorogovtsev2003evolution}, World Wide Web \cite{albert1999diameter, Albert2001, dorogovtsev2003evolution}, power-grid and industrial networks \cite{Albert2001, dorogovtsev2003evolution}, and transportation networks \cite{doi:10.1073/pnas.0407994102, LI2007693}. Dorogovtsev et al. and Albert et al. review numerous other additional examples of empirical scale-free networks in \cite{dorogovtsev2003evolution} and \cite{Albert2001}. Despite the physics-inspired character of the considered transport model, our work revealed several features that may find applications in real-world networks. First, the source-drain distance $\Delta$ is a simple yet useful indicator for network robustness when environmental changes randomly alter the links. Second, the inclusion of one additional link can further improve robustness, thus delaying the collapse. This work was focused on the collapse of transport, which is understood in terms of fluxes from the source to the drain. This a simple setup which focuses exclusively on a single pair of the source and drain nodes. We also included the condition (\textit{iii}) which stops the simulations when the main network is divided into two or multiple subgraphs (see Methods section 2.2). One source and drain and this breaking condition are reasonable for some applications and not relevant for others. For instance, in the transportation field, there can be multiple origins (sources) and destinations (drains) in one network. When an origin gets disconnected from the main network, transport remains possible, thanks to other origins. In a future work, we plan to extend this study and investigate the time of collapse in more general settings by removing condition (\textit{iii}) and including multiple sources and drains \cite{carmi2007transport, carmi2008transport}.

\section*{Conflict of interest declaration}
We declare we have no competing interests.

\section*{Funding}
The author(s) received no financial support for the research, authorship, and/or publication of this article.

\section*{Acknowledgments}
This article is dedicated to the work and commitment of Bernard Sapoval (1937--2020). D.S.G. acknowledges the Alexander von Humboldt Foundation for support within a Bessel Prize award.

\bibliographystyle{vancouver}
\bibliography{bibfile_CNetworks}

\end{document}